# *Refractive index retrieval of 3D printed materials for photonic applications*

*Joseph Arnold Riley[1†], Christian Johnson-Richards[1†], Noel Healy[1] and Victor Pacheco-Peña[1*]*

[1]*School of Mathematics, Statistics and Physics, Newcastle University, Newcastle Upon Tyne, NE1 7RU, United Kingdom*

**email:* victor.pacheco-pena@newcastle.ac.uk

*Abstract* – **The advent of additive manufacturing has opened opportunities to rapidly prototype devices and products ranging from automotive and aerospace applications to micro/nanoscale metastructures, as examples). Three-dimensional (3D) printing has become relevant for electromagnetic structures, integrated optics and photonics systems, however, the optical properties of commercially available 3D printed polymers at telecommunication wavelengths ($\lambda_0 = 1550$nm) is not always available. Provided the importance of 3D printing technologies, in this work, we evaluate both theoretically and experimentally the complex refractive index of four polymers including some recycled versions (namely Butenediol Vinyl Alcohol (BVOH), Polylactic Acid (PLA), Recycled Polyethylene Terephthalate (rPET), and recycled Polylactic Acid (rPLA)) as potential candidates for photonics applications. The 3D printed samples have thicknesses from $\sim 100$ to $400$ µm ($\sim 64\lambda_0$ to $\sim 258\lambda_0$, respectively). The experimental reflectance and transmittance spectra are extracted and used to retrieve the complex refractive index of each printed material demonstrating extinction coefficients in the order of $10^{-4}$ at $\lambda_0 = 1550$nm. The experimental results are validated using numerical simulations. Finally, as a proof-of-concept, a convex-planar lens and a Bragg mirror are designed and numerically evaluated, showing the potential of the proposed polymers for 3D printing photonic structures at telecommunication wavelengths.**

[†] These authors contributed equally to this work.

## 1. Introduction

Research and innovation to control wave-matter interactions has enabled the creation of new and improved applications including radars [1], [2], sensors [3], [4] and antennas [5]–[7], to name a few [8]–[14]. The fields of optics and photonics are constantly expanding, and their application landscape is vast with scenarios including sensors [15], [16], light generation [17], [18], topology [19]–[21], metamaterials and metasurfaces [22]–[26], computing with waves [8], [27]–[31] and spacetime systems [32]–[42], to name a few [43]–[46].

Additive manufacturing, in the form of three-dimensional (3D) pr{Citation}inting of thermal polymers, has evolved from being exclusively a method of prototyping into a precision manufacturing tool [47]–[50]. This has allowed the manufacture of complex designs not easily fabricated with traditional methods [49], [51]. For instance, 3D printed electromagnetic (EM) devices have been recently demonstrated at microwaves (such as lenses [52]–[54], antennas [55], waveguides [56]–[59] and metamaterials [60], [61]). As the printing resolution improved, applications at higher frequencies such as terahertz (THz) have also been recently reported [62], [63], showing the constant evolution of this important technology. When using 3D printing for the manufacture of structures working from microwaves up to the optical regime, the optical properties of the materials (such as the complex refractive index) are fundamental. In this context, selected 3D printed polymers have been characterised within microwave [64] and THz [65] spectral ranges. However, there has been limited characterisation of the complex refractive index of 3D printed polymers at near infrared wavelengths (~1550 nm). This may be due to multi-scattered light produced by fabricated samples at the printing resolution that is currently available [66], which has likely become a limiting factor in preventing the realisation of Fused Deposition Modelling constructed devices at the photonics scales.

Motivated by the importance of 3D printing and photonics technologies, in this manuscript we present our efforts to provide an initial complex refractive index library of some 3D printable polymers for the use in photonics applications at telecom wavelengths. The chosen 3D printable materials are BVOH, rPET, PLA and rPLA, as examples of water-soluble, degradable and recyclable materials. Multiple samples of different thicknesses for each of the chosen materials are fabricated using a commercially available 3D printer. The transmittance, $T$, and reflectance, $R$, spectra are experimentally measured at telecom wavelengths (1520 to 1630 nm) for each of the fabricated samples and implemented to extract their corresponding complex refractive index using an analytical method that takes into account the multiple scattering produced by the samples [67], [68]. Finally, two photonics devices (namely a converging lens and a Bragg mirror) are designed and numerically evaluated, demonstrating the potential of these polymers for photonics applications.

## 2. Design and materials

In this section, the design of the samples, 3D printing method and characterisation procedure used to retrieve the complex refractive index is presented.

### 2.1 Fabrication

The polymers were 3D printed using a commercial Raise3D pro2 printer (via Fused Deposition Modelling (FDM)) [50] fitted with a 0.2 mm nozzle attachment. A glass-based hot bed was retrofitted, chosen to minimise the roughness of the printed sample on its underside surface. Each sample was designed to have a region of interest (ROI, defined as the illuminated region by the incident beam) with dimensions of 15 mm × 15 mm. The 3D printer was configured to use a 100% rectilinear type infill pattern with a target layer height of 0.025 mm to produce a solid ROI [69] and a print speed of 15 mm/s. To improve the rigidity of the sample and smoothness of the ROI, a 2 mm wide and 1.5 mm thick frame was added to the design (i.e., this frame is made out of the same material as the sample). In so doing, the overall size of the sample was 19 mm × 19 mm. A rendered schematic representation of the samples is shown in Fig. 1A(i) where the ROI and the frame are highlighted in red and grey, respectively.

The materials (BVOH, rPET, PLA and rPLA) were sourced from commercial suppliers in spools of filament with 1.75 mm diameter. Specifically, BVOH and PLA filaments were purchased from UltraFuse™ and Raise3D™, respectively, while rPET and rPLA filaments are both manufactured by Filamentive™. To reduce the effect of additives on the EM properties of the materials, the polymers were selected with no dye and, when this was not possible, white colour was selected. Samples of each polymer were printed with multiple thickness between 100 μm to 400 μm ($\approx 64.52\lambda_0$ and $258.06\lambda_0$ where $\lambda_0 = 1550$ nm). As shown in Fig. 1A(i), three regions are marked as A-C. The position of these three regions is as follow: position B is placed at the centre of the sample while A and C are located 3 mm away from B. These positions are used to measure the thickness of the fabricated sample, a parameter to be used later in the process to retrieve the complex refractive index. As it will be shown, the printed thickness deviated from the designed thickness. This may be due to the expansion and cooling of the filaments and small deviations in the calibration of the print bed. However, the experimentally measured values of each sample were used in the calculation of the optical properties of the samples.

As examples, pictures of the fabricated samples for a designed thickness of $\approx 200$ μm ($129.03\lambda_0$) are shown in Fig. 1A(ii-v) for BVOH, rPET, PLA and rPLA respectively. Microscopic images for the samples are also shown as insets for completeness. These images are taken from position A (Fig. 1A(i). As observed, the line pattern used to print the samples is visible on their surface, with a solid surface and minimal air gaps.

### 2.2 Experimental Setup

To characterise the printed polymers, an open-bench setup was used to measure the $R$ and $T$ spectra (see the schematic representation of the setup in Fig. 1B along with a top-view of the real setup in Fig. 1C). In this setup, a continuous wave Tunics T100S-CL 1520-1630 nm tuneable laser is routed via a Corning single mode fibre (SMF-28, shown in Fig. 1B,C(i)) to a Thorlabs F280FC-1550 beam collimator Fig. 1B,C(ii) with a Full-width at Half-Maximum (FWHM) = 3.6 mm. To measure $T$, a photodetector (Fig. 1B(v)) was positioned behind the sample holder, itself represented by the red block in Fig. 1B and Fig. 1C. Simultaneously, another photodetector (Fig. 1B,C(vii)) was used to measure $R$ via a 55:45 beamsplitter (Fig. 1B,C(iv)). Any unwanted reflections, particularly from the photodetectors, were re-directed to a beam absorber Fig. 1B,C(xi). The sample was placed in a bespoke holder (with the smooth underside surface facing in the $-z$ direction), allowing the sample to be rotated about the $x$- and $y$-axis. To align each beam with the two photodetectors, adjustable irises (Fig. 1B,C(iii, vi)) were used. The sample holder was fixed to a manual linear stage aligned to the $x$-axis allowing the sample to be translated and illuminate positions A-C (Fig. 1A(i)).

Prior to measuring each sample, to enable subsequent normalization of the sample spectra, reference spectra for the reflectance were measured using a fused silica broadband reflective mirror, while an open optical path (no sample) was used for the transmittance. The sample was then mounted in the holder, and the optical path was carefully aligned to ensure normal incidence of the illumination beam. For reflection, alignment was achieved by closing the iris in Fig. 1B(vi) and adjusting the sample holder rotation until the power was maximised at the reflectance photodetector. The spectra were collected sequentially from three positions (A-C from Fig. 1A(i)) across the sample to account for spatial variability of the print. At each position, the printed sample was translated into place, the normal reflection was re-aligned, and both reflectance and transmittance spectra were recorded. All sample spectra were subsequently normalized to the reference spectra.

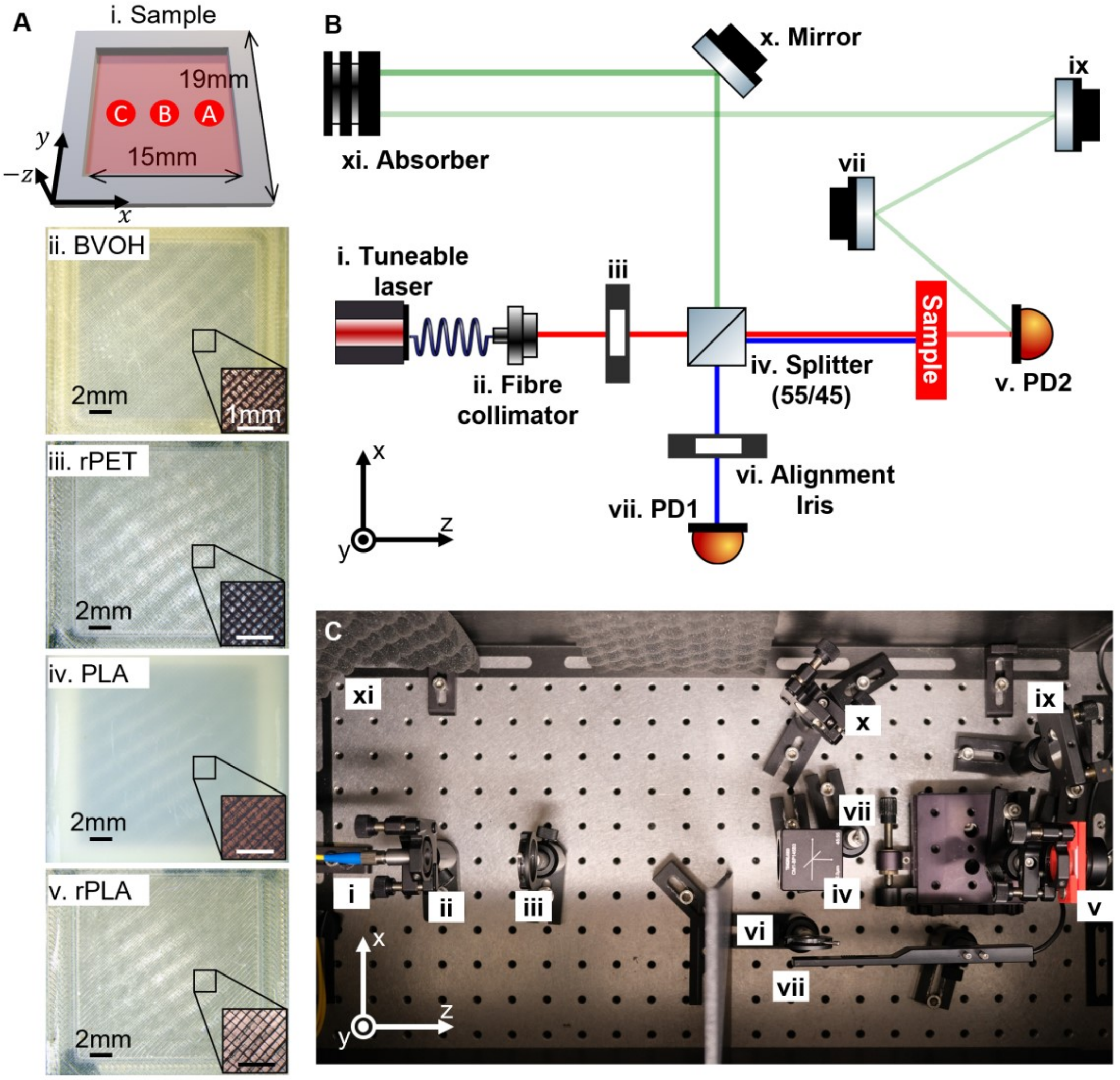

**Figure 1 Experimental setup A(i)** Schematic representation of the sample and **(ii-v)** macroscope (microscope) image (inset) of the BVOH, rPET PLA and rPLA samples under test showing the $-z$ interface. **B** top-down schematic of the experimental characterisation, **C** top-down photograph of the experimental setup used to characterise the samples shown in **A**.

## 3. Results and Discussion

### 3.1 Description of the experimental and simulation process

The schematic representation of the process followed in this work is presented in Fig. 2. After fabricating the samples, they were experimentally characterized in free space (see left panel from Fig. 2) by measuring their corresponding $T$ and $R$ spectra at telecom wavelengths (specifically between 1520 nm and 1630 nm with a spectral resolution of 1 nm). The $T$ and $R$ spectra (top panel from Fig. 2) were then used to calculate the real ($n$) and imaginary ($\kappa$, extinction coefficient) values of the complex refractive index for each printed material using an incoherent interference method [67], [68].

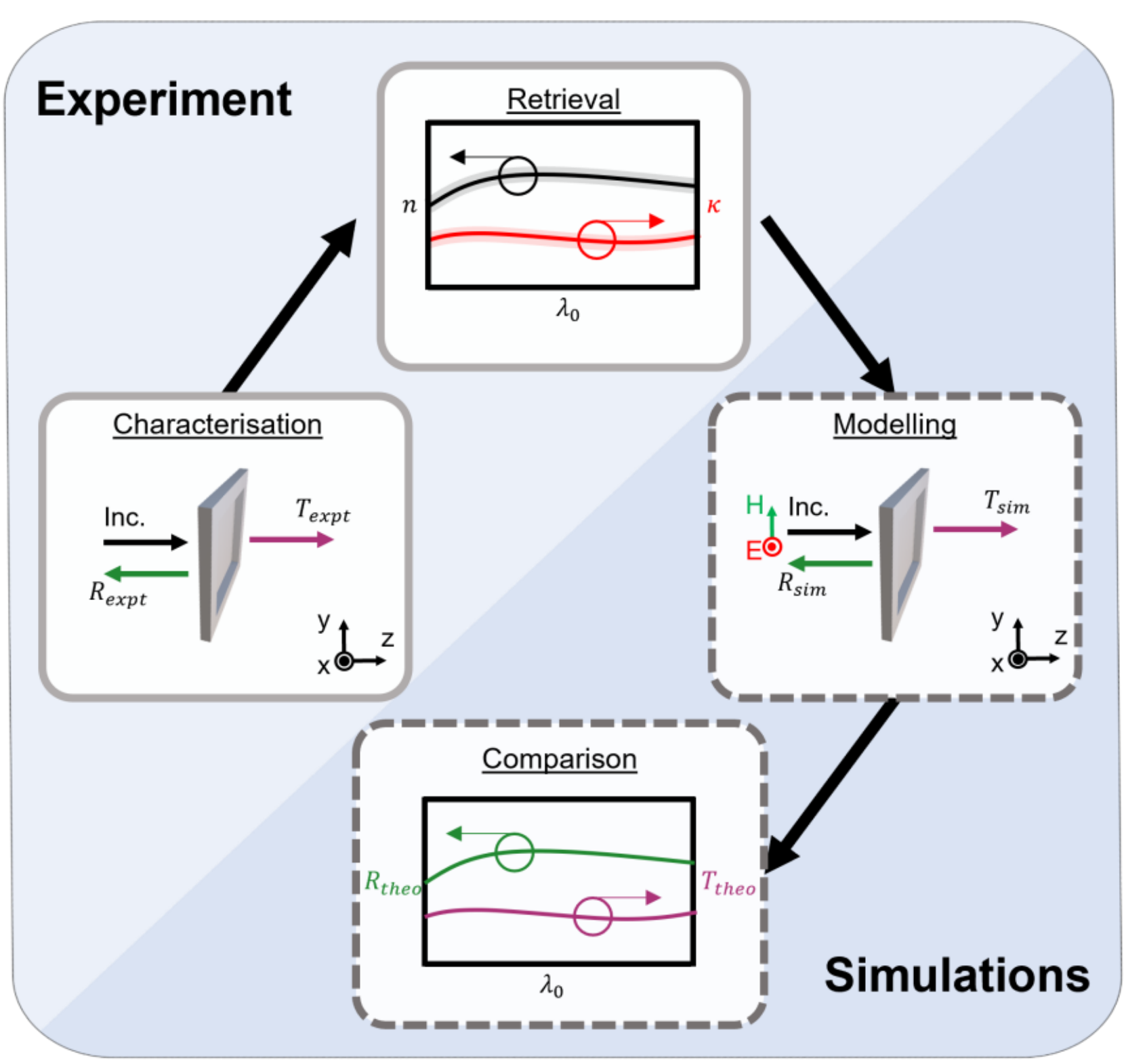

**Figure 2| Schematic representation of the 3D printed polymer characterisation steps.** Starting with the experimental section (light blue background) with reflectance (green) and transmittance (purple) characterisation followed by the retrieval of the $n$ and $\kappa$ values. These results are then used in the simulation section (dark blue background) where the samples are modelled as blocks to reproduce the reflectance and transmission spectra. At this point the experimental, analytical and numerical results are compared.

To validate the experimental results, numerical simulations were carried using the open source MEEP finite time domain solver (right panel from Fig. 2) [70]. To do this, the samples for each material were modelled using blocks illuminated by a planewave within the spectral range under study. The extracted experimental values of $n$ and $\kappa$ for each printed material were used in the numerical calculations. With this configuration, the numerical results for the $R$ and $T$ spectra were compared with analytical calculations using the incoherent interference method and the experimental results (bottom panel from Fig. 2, see details below). This process allowed us to provide a full experimental and numerical validation of the calculated complex refractive index for each material, as it will be shown below.

### 3.2 Transmittance and Reflectance Spectra

Using the experimental setup described in Fig. 1, the resulting $R$ and $T$ measured spectra are presented in Fig. 3. The results are organised in rows by material (from top to bottom: BVOH, rPET, PLA and rPLA) and in columns by thickness group (with thicknesses raging from 100 μm to 400 μm). Given the constraints of the spatial resolution of the available 3D printer, it is expected that the fabricated samples may vary in thickness (as explained above). To account for this, and to reduce variation, the $R$ and $T$ spectra for each printed material were measured for multiple samples of the same target thickness. Specifically, three samples of the same material and thickness were measured to obtain three different $R$ and $T$. These results are then presented as blue, green and pink plots in Fig. 3. For instance, looking at the results for BVOH (first row from Fig. 3), the panels from left to right show the results for a targeted thickness of ∼200 μm, ∼300 μm, ∼400 μm, respectively. Moreover, the results in Fig. 3 represent the obtained datapoints with their corresponding errors (each datapoint in Fig. 3 is calculated from the average of 10 measurements repeated 3 times at positions A-C in sequence for a total of 90 datapoints per wavelength per sample).

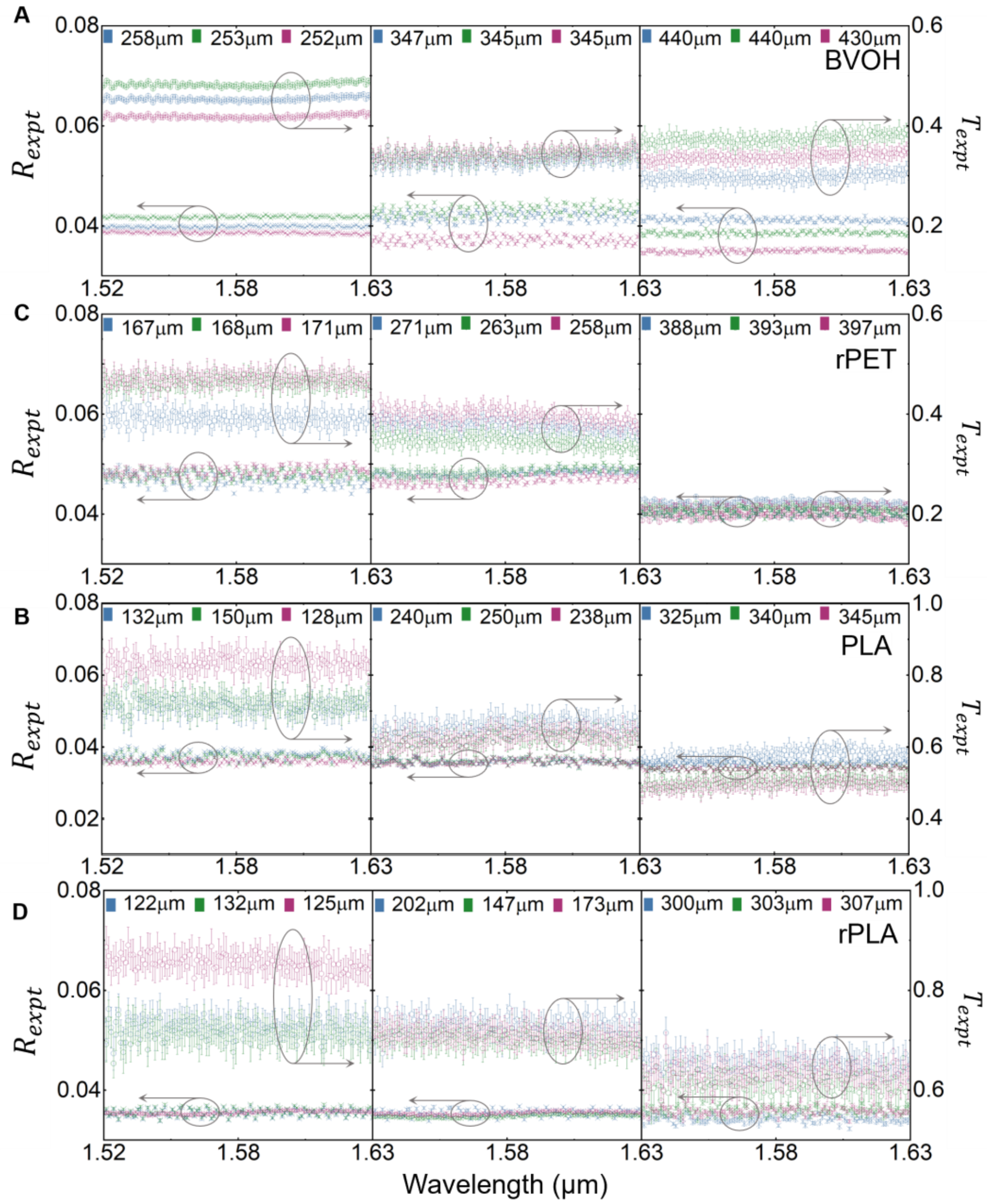

**Figure 3| Experimentally measured transmittance and reflectance spectra. A-D** organised in rows by material (BVOH, rPET, PLA, rPLA, respectively) and columns by sample thickness. Three repeats of each material and thickness are included in the colours blue, green and pink, respectively. Each datapoint includes an error bar.

As observed in Fig. 3, for all the samples, $T$ is reduced when increasing the thickness of all the samples. This is an expected result as these are lossy dielectrics. Interestingly, one can notice that the error bars for the measured $R$ spectra are smaller compared to those obtained for $T$. This may be explained by the manufacturing technique: as explained above, the blocks were 3D printed on a glass bed. The surface printed on glass (called the ROI interface in the $-z$ direction) was illuminated by the incident beam. Due to this, this first printed surface was smoother compared to that of the last printed surface ($+z$ surface), meaning that the latter retained higher surface roughness from the print nozzle. This is the reason why the reflected signal remains mostly specular while scattering is more evident for the transmitted signal. To account for this limitation, the photodetector for $T$ (Fig. 1.B,C(v)) was placed close to the $+z$ ROI surface ($\approx$ 12.4 mm from the $-z$ interface of each sample to photodiode surface) to capture the scattered signals. Moreover, note that air cavities may exist within the sample which would impact both $T$ and $R$. However, as it will be shown below, the extracted complex refractive index for all the samples is consistent between samples meaning that their effect may be minimal in our case. The rough quality of the $+z$ surfaces is expected to be the most limiting factor of the results presented in this work as the measured value of $T$ may be lower to that compared of a bulk material with a smooth output surface. This could introduce an overestimation of values of $\kappa$. However, even with this drawback, this overestimation will better represent a realistic FDM printed sample when constructed with the current standard of FDM printing technology.

### 3.3 Retrieved $n$ and $\kappa$ spectra

The $n$ and $\kappa$ spectra of the printed materials are calculated using the analytical method described in [67], [68] and outlined in the methods section for completeness. In this method, the scattering produced at the output of the samples is taken into account. With this in mind, the $n$ and $\kappa$ spectra for the measurements described in Fig. 3 are shown in Fig. 4 as blue and red datapoints with error bars, respectively. The mean of the extracted values of the three samples for each of the three design thicknesses was also calculated and the results are shown as black lines in Fig. 4.

As observed, the samples present a flat variation of $n$ across the measured spectra, meaning that we are working away from any resonances in the material (for PET this is expected outside the measured wavelength range [71]). When looking at the results at the specific telecom wavelength of $\lambda_0 = 1550$ nm, the real component of refractive index of BVOH , rPET, PLA and rPLA is $1.4542 \pm 0.0267$, $1.4970 \pm 0.0168$, $1.3736 \pm 0.0160$, and $1.3586 \pm\ 0.0146$, respectively. Due to the limited studies on these materials within this wavelength range, a direct comparison for all materials is not possible, however, the values for rPET in our work are in line with [71], [72] for PET. Additionally, the real component of the refractive index of PLA and rPLA is approximately equal, suggesting that recycled versions of PLA can provide similar optical response as the original version. The extinction coefficient for each material at $\lambda_0$ is BVOH $3.3576 \times\ 10^{-4} \pm 4.4703 \times 10^{-5}$, rPET $4.7612 \times 10^{-4} \pm 6.0144 \times 10^{-5}$, PLA $2.1032 \times\ 10^{-4} \pm 3.3323 \times 10^{-5}$ and rPLA $2.0498 \times 10^{-4} \pm 5.9023 \times 10^{-5}$. The result for rPET is slightly higher than similar results found in [72] for PET. This is potentially due to the limitation in the sample manufacturing mentioned earlier where, due to high $+z$ interface surface roughness, all the scattered signal in transmission may not be captured in its entirety by the detector. Again, comparing PLA to rPLA there is not a significant difference in the values of $\kappa$ demonstrating that the recycled version is a good option for photonic structures at telecom wavelengths.

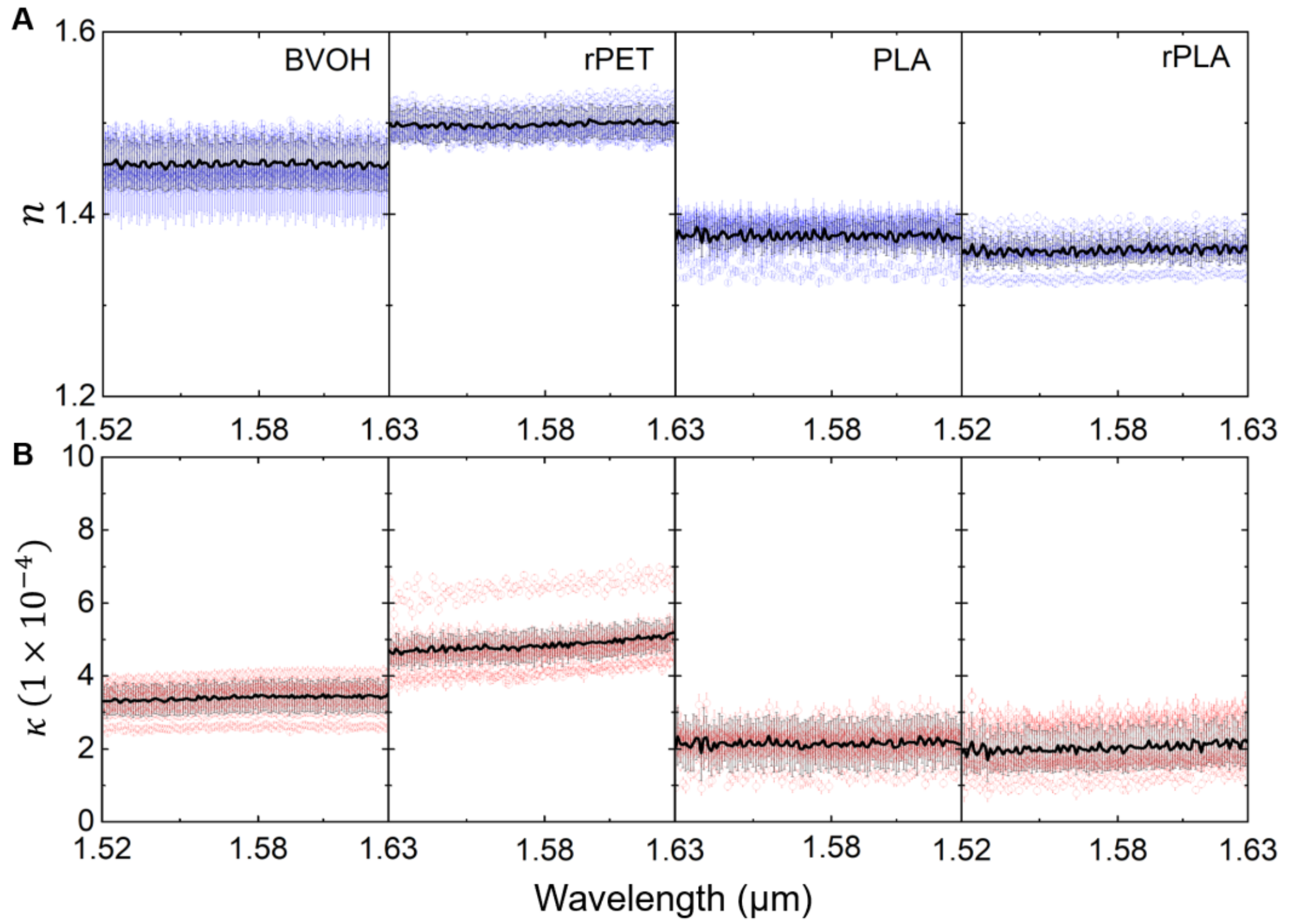

**Figure 4|** Retrieved $n$ in panel A (blue) and $\kappa$ in panel B (red) of the printed samples organised into columns by material (from theft to right: BVOH, rPET, PLA, rPLA). The coloured datapoints represent the retrieved $n$ (blue) and $\kappa$ (red) for all measurements presented in Fig. 3. The black lines represent the mean values of $n$ and $\kappa$.

### 3.4 Numerical Comparison

Following the diagram presented in Fig. 2, analytic and numerical studies of the $R$ and $T$ spectra were carried out to verify the extracted complex refractive index for all materials under study. The model is shown schematically in Fig. 5A and consists of an ideal block of polymer with complex refractive index $\tilde{n}$ and thickness $t$, immersed in air. The optical properties for each material were modelled using a sum of Lorentzian fit with the experimentally retrieved mean $n$ and $\kappa$ shown in Fig. 4. The fits for each material are shown in Fig. 5B as blue and red solid lines for $n$ and $\kappa$, respectively. Select sample thicknesses of 253, 168, 132 and 132 μm for BVOH, rPET, PLA and rPLA, respectively, were used in the following studies. To fully compare all the results, three methods were used to calculate the $R$ and $T$ spectra: i) an incoherent interference method (Eq. 5-6 from the methods section, which were also used to retrieve the experimental values from Fig. 4), ABCD method (Eqn. 7-8 from the methods) and numerical simulations (exact numerical setup is detailed in the methods section).

With this configuration, the results of the $R(T)$ spectra shown in Fig. 5C as follows: solid dark green(purple), light green(purple) and dashed dark green(purple) lines for the incoherent interference, ABCD and numerical method, respectively. From these results, it can be observed that Fabry-Perot (FP) resonances are obtained for the ABCD and numerical method while smooth spectra are obtained for the incoherent interference method. These results are as expected as the incoherent interference method considers a block of material with a thickness and surface roughness much larger than the operating wavelength (i.e., no interference will occur as there are no coherent reflections) [67], [68].

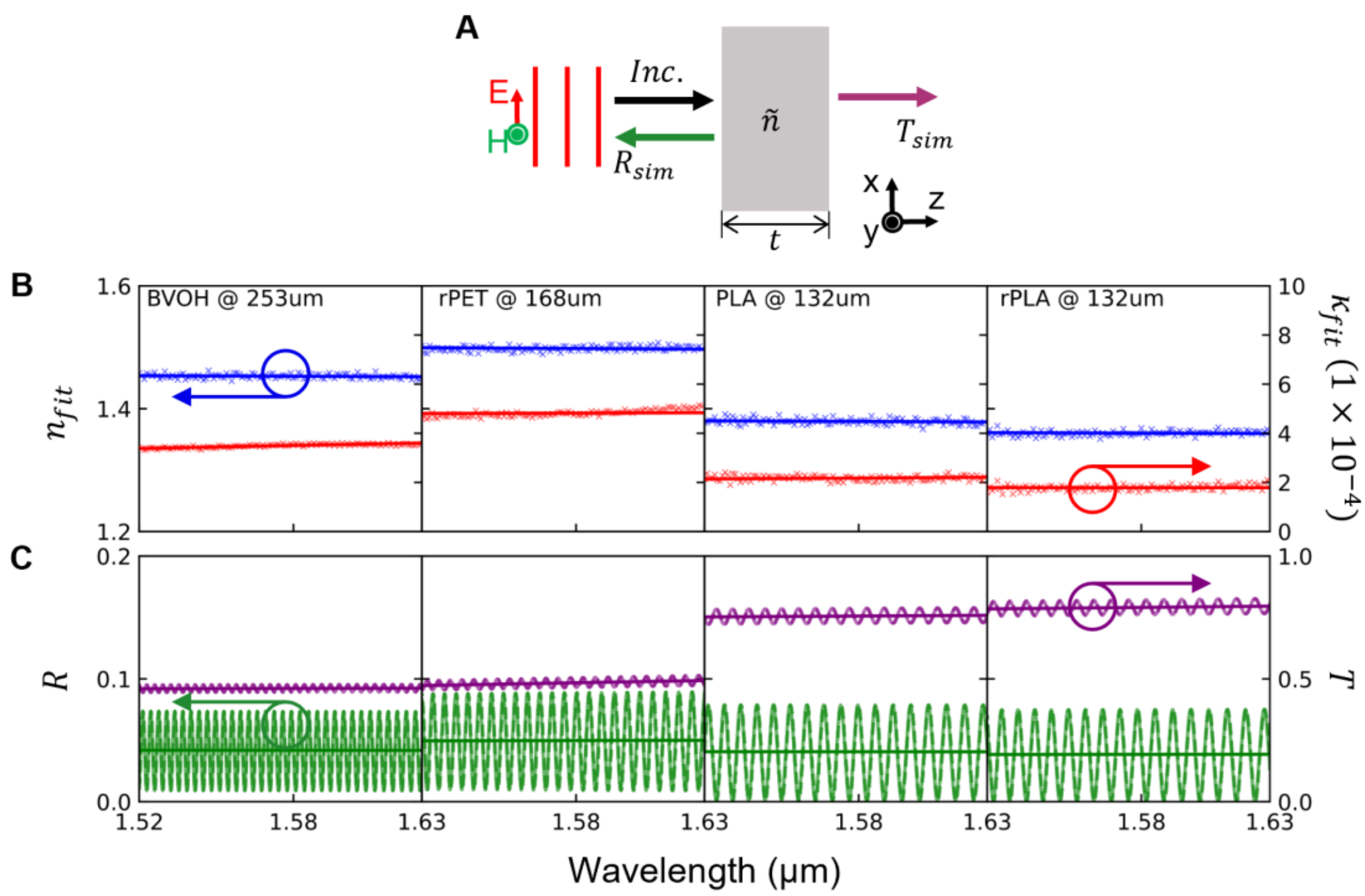

**Figure 5| A** Schematic representation of the numerical/theoretical model used to calculate the transmittance and reflectance spectra. **B** Lorentzian fitted $n$ and $\kappa$ for each material (solid lines) along with the experimental data from Fig. 4. **C** Analytical results of the extracted $R$ and $T$ spectra using the incoherent interference method (dark green and dark purple lines, respectively) and the ABCD method (light green and purple, respectively) for each material (BVOH thickness 253 μm, rPET thickness 168 μm, PLA thickness 132 μm and rPLA thickness 132 μm). The numerical simulation results are also shown in this figure as dashed lines.

Moreover, note how a significant variation of the $R$ and $T$ spectra is present for the ABCD and numerical results (due to FP resonances), however, the results are approximately centred on those produced by the incoherent interference method. The latter results are in agreement with the experimental $R$ and $T$ data presented above in Fig. 3. Taking into account the results from Fig. 5C, the incoherent interference method likely represents the experimental setup more closely due to a rough second interface present on the printed samples. Therefore, we can directly compare the numerically calculated and experimental values of for $R$ and $T$: for example, at $\lambda_0 = 1550$nm the calculated value of $R$ using the incoherent interference method for BVOH, rPET, PLA and rPLA is $4.19 \times 10^{-2} \pm 0.43 \times 10^{-2}$, $4.97 \times 10^{-2} \pm 0.31 \times 10^{-2}$, $4.07 \times 10^{-2} \pm 0.30 \times 10^{-2}$ and $3.85 \times 10^{-2} \pm 0.31 \times 10^{-2}$, respectively, which align well with the experimentally extracted values ($4.13 \times 10^{-2} \pm 7.48 \times 10^{-5}$, $4.80 \times 10^{-2} \pm 3.80 \times 10^{-4}$, $3.73 \times 10^{-2} \pm 4.45 \times$

$10^{-4}$, $3.47 \times 10^{-2} \pm 2.38 \times 10^{-4}$, respectively).

**3.5 Applications**

For completeness, here we present two final studies to explore the potential of the characterized materials for photonics applications. Specifically, we present the design and numerical results of a focusing lens and a Bragg mirror as relevant elements in optics and photonics. The ability to design and fabricate such elements using the proposed materials may open opportunities of rapid manufacture and reduce costs of prototyping. To begin with, the schematic representation of a focusing lens along with the numerical simulations using COMSOL Multiphysics® are shown in Fig. 6A. Here, the lens was designed at $\lambda_0 = 1550$ nm using the known lens maker equation which relates the effective focal length ($f$) to the refractive index of the lens ($n_2$) and the surrounding medium ($n_1$, free space in our case) as well as the surface curvatures [22], [73] as $f^{-1} = (n_2 - n_1)(n_1)^{-1}\left(\frac{1}{R_1} - \frac{1}{R_2}\right)$; with $R_1$ and $R_2$ as the radii of curvature of the illuminated and back faces, respectively. A convex-planar profile ($R_2 = \infty$) was selected following [22]. The lens material was chosen to be rPLA ($\widetilde{n_2} = 1.36 + i2.0 \times 10^{-4}$) as an example. The lens was designed with a thickness, $d = \lambda_0$ and $R_1 = 2\lambda_0$ to produce a focus at a distance of $\approx 5.56\lambda_0$ from the principal plane of the lens (or at a distance of $\approx 4.82\lambda_0$ from the back face of the lens[73]). The designed lens was then illuminated by an $x$-polarized planewave.

The numerical results of the power enhancement (defined as the ratio of the power distribution with and without the lens) on the $xz$-plane is shown in Fig. 6A(ii) demonstrating that a focus is produced. For completeness, the power enhancement along the $z$-axis for $x = y = 0$ is shown on the left panel from Fig. 6A(iii). Here, a power enhancement of $\approx 6.77$ can be observed at $z = 2.47\lambda_0$ from the output surface ($z = 0$) or $3.13\lambda_0$ from the principal plane of the lens. The realized focal length being closer to the lens is likely a result of the inclusion of losses which are not accounted for in the lens maker equation. To evaluate the spatial resolution of the lens, the power enhancement along the $x$- and $y$- axis at the position of the focus is shown in the right panel from Fig. 6A(iii). From these results, the Full-Width at Half-Maximum is $FWHM_{x,y} \approx 1.04\lambda_0$ noting that such large value is expected as the idea here is to show the design of a lens using the characterized materials (the improvement of the spatial resolution of the lens is outside of the scope of this work). Finally, to better observe the performance of the lens, we provide the power enhancement on the *xy*- plane at three different positions along the $z$-axis, namely $z = \lambda_0$, $2.47\lambda_0$ and $5\lambda_0$, i.e. below, at the focal length, and away from it. Despite the fact that further advances will be required to improve the surface quality on the final layers (outside of the scope of this work), these results demonstrate that these materials may be used for lenses at telecom wavelengths.

The second study focused on the feasibility of using the studied polymers to design a Bragg mirror [74], [75]. A schematic representation of the Bragg Mirror is shown in Fig. 6B(i) where a periodic structure of alternating layers of two different materials (rPET and air, with thicknesses of $5.1\lambda_0$ and $4.85\lambda_0$, respectively) is presented. A total of 80 layers were implemented. Using these values, the reflection and transmission spectra were calculated using the ABCD matrix method [76] and the results are shown in Fig. 6B(ii). From these results, the characteristic spectra of a Bragg mirror can be observed, showing a high, near unity, reflection with low transmission around the designed wavelength of $\lambda_0 = 1550$ μm. To further corroborate the design, numerical simulations were carried out using COMSOL Multiphysics® and the results of the transmittance and reflectance are shown in Fig. 6B(iii). These results corroborate a minimum of transmission at a central wavelength of ≈ 1535.22 nm (≈$0.99\lambda_0$). This central wavelength could be shifted to fall exactly at the designed value by numerically optimizing the thickness of the layers, however this is outside of the scope of this work. Importantly, note how for the spectral bands where high transmission (lower reflection) is expected, both transmission and reflection are low in the numerical simulations. This is because of the effect of material losses of the polymer, resulting in a significant proportion of the incident wave being absorbed within the structure rather than transmitted. A similar performance occurs within the spectral region where high reflection is expected, noting how reflection is not unitary. These results are as expected as material losses are included in the numerical simulations while the theoretical calculations from Fig. 5B(ii) assume zero losses.

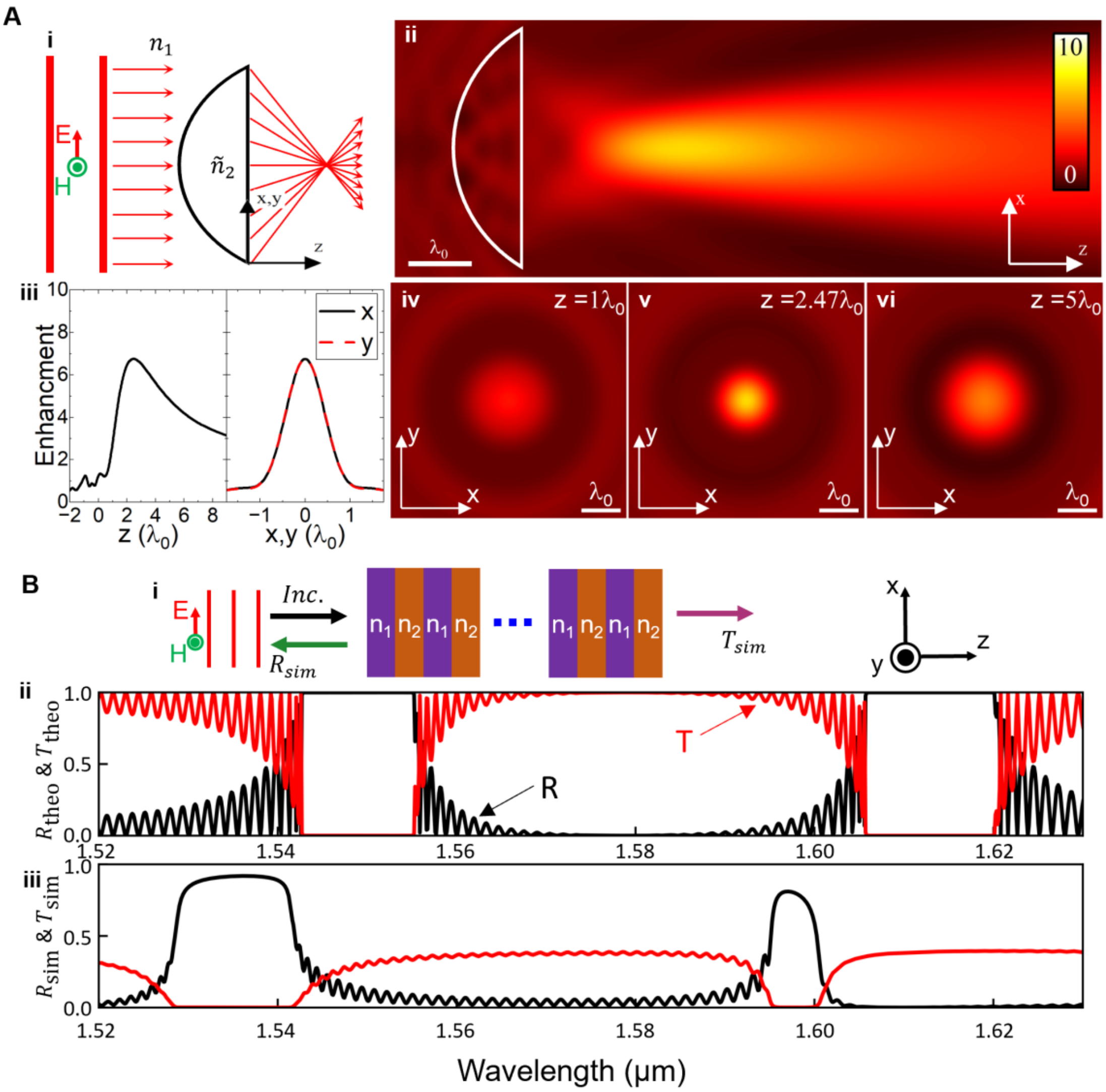

**Figure 6| Schematic representations and numerical results of a 3D convex-planar lens and a Bragg mirror. A** (i) schematic of a 3D convex-planar lens ($R_1 = 2\lambda_0$) made from a rPLA and immersed in free space. Numerical results of the power enhancement: (ii) on the $xz$-plane, (iii) along the direction of propagation ($z$-axis) at $x = y = 0$ (left) and along the $x$ (black) and $y$ (red) transversal axes at the position of the focus (right), and in the $xy$- plane at $z = 1\lambda_0$ (iv), $z = 2.47\lambda_0$ (v, focal length), and $z = 5\lambda_0$ (vi). **B** (i) Schematic representation of a Bragg mirror consisting of alternating layers of two materials ($n_1$, rPET, and $n_2$, air) with thicknesses of $t_1, t_2 = 5.1\lambda_0, 4.85\lambda_0$ respectively. Analytical (ii) and numerical (iii) results of the reflectance (black) and transmittance (red) spectra of the Bragg mirror.

## 4. Conclusions

In this work, the optical properties (complex refractive index) of four 3D printable materials (BVOH, rPET, PLA and rPLA) were characterized within the spectral range of 1520 nm and 1630 nm. For this, FDM printing was implemented to fabricate the samples using different thicknesses ($\sim 100 - 400$ µm). To validate the experimental findings, numerical simulations and analytical calculations were performed, demonstrating an agreement between the results. Finally, the design and numerical results of a converging lens and a Bragg mirror using some of the characterized materials were presented, highlighting the potential of 3D printed polymers for photonics applications.

## 5. Methods

### 5.1 Sample Manufacture

Each polymer required different print temperatures and conditions as summarized in Table 1.

**Table 1| Printing parameters and filament characteristics for each polymer used in this study**.

| **Material** | **Colour** | **Temperature (◦C)** | |
|---|---|---|---|
| | | **Bed** | **Nozzle** |
| BVOH | Natural | 63 | 205 |
| rPET | Transparent | 70 | 210 |
| PLA | White | 60 | 205 |
| rPLA | Natural | 60 | 205 |

### 5.2 Complex refractive index extraction and ABCD method

The complex refractive index ($n + i\kappa$) was determined using[67], [68]. Full details can be found in [67], [68], but we summarize them here for completeness. As it is known, when considering a dielectric slab of thickness $\Delta z$, the amplitude of the wave will decay by $e^{\frac{-\alpha\Delta z}{2}}$ (where $\alpha = \frac{4\pi\kappa}{\lambda}$, and $\lambda$ is the wavelength of the incident signal) and its phase will change as $e^{i\Delta\phi} = e^{\frac{i\delta}{2}}$ with $\Delta\phi = \beta\Delta z$ ($\beta$ as the propagation constant and $\delta = \frac{4\pi n\Delta z}{\lambda}$). The total reflectance $R_{tot}$ and transmittance $T_{tot}$ from the multiple reflections can be evaluated as [67], [68]:

$$R_{tot}(\delta) = \frac{\rho\left(1 - 2e^{-\alpha\Delta z}\cos\delta + e^{-2\alpha\Delta z}\right)}{1 - 2\rho e^{-\alpha\Delta z}\cos\delta + \rho^2 e^{-2\alpha\Delta z}} \tag{1}$$

$$T_{tot}(\delta) = \frac{(1-\rho)^2 e^{-\alpha\Delta z}}{1 - 2\rho e^{-\alpha\Delta z}\cos\delta + \rho^2 e^{-2\alpha\Delta z}} \tag{2}$$

where $\rho$ denotes the single-interface reflectance at the free-space/polymer (dielectric) boundary. As the incident illumination was effectively incoherent over the sample thickness due to printing tolerances over the sample, the observed reflectance ($R_{bulk}$) and transmittance ($T_{bulk}$) were obtained by averaging the phase-dependent expressions over all possible phase shifts ($\delta$):

$$R_{bulk} = \left(\frac{1}{2\pi}\right)\int_0^{2\pi} R_{tot}(\delta)d\delta \tag{3}$$

$$T_{bulk} = \left(\frac{1}{2\pi}\right)\int_0^{2\pi} T_{tot}(\delta)d\delta \tag{4}$$

Evaluating these integrals yields closed-form expressions for the reflectance and transmittance at normal incidence (see [67], [68] for further details):

$$R = \frac{\left[\rho\left(1 + (1-2\rho)e^{-2\alpha\Delta z}\right)\right]}{\left[1 - \rho^2 e^{-2\alpha\Delta z}\right]} \tag{5}$$

$$T = \frac{\left[(1-\rho)^2 e^{-\alpha\Delta z}\right]}{\left[1 - \rho^2 e^{-2\alpha\Delta z}\right]} \tag{6}$$

The extracted $n$ and $\kappa$ values for each material were obtained by substituting the measured $R$ and $T$ spectra from Fig. 3. For the ABCD method used in section 3.4 the following equations were used to extract $R$ and $T$ [76]:

$$R = \left|\frac{(\tilde{n}^2 - 1)\left(-1 + e^{i2\tilde{n}k_0\Delta z}\right)}{(\tilde{n}+1)^2 - (n-1)^2 e^{i2\tilde{n}k_0\Delta z}}\right|^2 \tag{7}$$

$$T = \left|\frac{4\tilde{n}e^{i\tilde{n}k_0\Delta z}}{(\tilde{n}+1)^2 - (n-1)^2 e^{i2\tilde{n}k_0\Delta z}}\right|^2 \tag{8}$$

where $\tilde{n}$ is the complex refractive index and $k_0$ is the free space wavenumber.

### 5.3 Numerical Models

The numerical simulations in Fig. 5 were performed using an open-source finite-difference (FDTD) time-domain software package: MEEP [70]. In this model a one-dimensional system was considered with a mesh density of 100 pixels/μm and PML layers applied in the $\pm z$ direction with a thickness of 50μm. A pulsed source was used and centred at a wavelength of $\approx$ 1573 nm with a spectral pulse FWHM of $\approx$ 52.7 nm. The structures were illuminated by a planewave with the electric field along the $y$- axis (perpendicular to the propagation direction $+z$ direction). The $R$ and $T$ spectra were calculated using the built-in flux region monitors placed at the $\pm z$ PML boundaries and normalised against an identical model with the dielectric block being removed. For the Lorentzian fits used to model the dispersive materials, 10 Lorentzian were used for BVOH and rPET and 2 Lorentzian for PLA and rPLA.

To model the devices shown in Fig. 6, COMSOL Multiphysics® was used with a general physics "Extremely Fine" mesh preset and one whole geometry mesh refinement step. The lens was immersed in free space with scattering boundary conditions in the $\pm x$, $\pm y$ and $+z$ directions. An incident planewave was applied from the $-z$ boundary of the simulation box using a scattering boundary condition. A 2D model was used to calculate the results for the Bragg mirror shown in Fig. 6B. Perfectly electric conductor boundary conditions were used on the $\pm y$ boundaries with planewave illumination applied by an input port on the $-x$ boundary.

## 6. Acknowledgements

This work was supported by the Leverhulme Trust under the Leverhulme Trust Research Project Grant scheme (RPG-2023-024) and the Royal Society under the Research Grant scheme (RGS\R2\212234). V.P-P. and J.A.R would like to thank the support from the Engineering and Physical Sciences Research Council (EPSRC) under the EPSRC DTP PhD scheme

(EP/R51309X/1). 

## 7. Conflicts of interests

The authors declare no conflicts of interests.

## 8. Data Accessibility Statement

The data that support the findings of this study are available from the corresponding author upon reasonable request.